# Combined Bulk and Surface Radiation Damage Effects at Very High Fluences in Silicon Detectors: Measurements and TCAD Simulations

F. Moscatelli, D. Passeri, *Member, IEEE,* A. Morozzi, *Student Member, IEEE,*
Roberto Mendicino, *Student Member, IEEE,* G.-F. Dalla Betta, *Senior Member, IEEE,* G.M. Bilei

*Abstract*– In this work we propose a new combined TCAD radiation damage modelling scheme, featuring both bulk and surface radiation damage effects, for the analysis of silicon detectors aimed at the High Luminosity LHC. In particular, a surface damage model has been developed by introducing the relevant parameters ($N_{OX}$, $N_{IT}$) extracted from experimental measurements carried out on p-type substrate test structures after gamma irradiations at doses in the range 10-500 Mrad(Si). An extended bulk model, by considering impact ionization and deep-level cross-sections variation, was included as well. The model has been validated through the comparison of the simulation findings with experimental measurements carried out at very high fluences ($2\times10^{16}$ 1 MeV equivalent n/cm$^2$) thus fostering the application of this TCAD approach for the design and optimization of the new generation of silicon detectors to be used in future HEP experiments.

## I. Introduction

COMPREHENSIVE TCAD models of radiation induced damage effects in solid-state devices have been developed to better understand the behavior of silicon detectors during operations at the Large Hadron Collider [1-6]. In particular, bulk damage impacts on the detector operation by introducing acceptor- and donor-type deep-levels acting as generation/recombination centers and/or trap states. This leads to higher leakage current, change in the effective space charge concentration and decrease of the charge collection efficiency due to charge trapping. On the other hand, surface damage introduces oxide charges and interface traps, which strongly influence the break-down voltage, the inter-electrode isolation and capacitance, and might also impact the charge collection properties of silicon sensors.
Our group previously developed two three-level models (usually referred as "Perugia" model), one for n-type [1] and one for p-type substrate [2], able to reproduce the radiation damage macroscopic effects up to fluences of the order of $10^{15}$ n/cm$^2$ 1MeV equivalent neutrons. The much higher fluences, at least one order of magnitude higher, expected at the High Luminosity LHC (HL-LHC) impose new challenges for the design of effective, radiation resistant detectors. To this purpose, TCAD tools should be proficiently adopted to evaluate different technological and design options in advance. However, the extension of the past model is not straightforward. New effects (e.g., avalanche multiplication and capture cross section dependencies on temperature) have to be taken into account, at the same time keeping the solid physically based approach of the modeling scheme. In order to create as comprehensive model as possible, we relied on both measured parameters of deep- level states (e.g. position within the gap, cross-sections...) and experimentally-measured oxide charge and interface traps. We also avoided fitting parameters to preserve the generality of the approach. In particular, the voltage stability and the charge-collection properties of segmented silicon sensors are strongly influenced by the charge layers and potential distribution at the sensor surface, the charge density within the oxide and passivation layers, and by Si-SiO$_2$ interface traps. The addition of these features to the modelling scheme will allow its application to the optimization of different detectors, eventually provided by different vendors.

Measurements on dedicated test structures and TCAD device-level simulations have been carried out aiming at a better understanding of complex phenomena related to the interplay of bulk, surface and interface radiation damage effects. We therefore focused our analyses on the properties of SiO$_2$ layer and of Si-SiO$_2$ interface using gate-controlled diodes (gated diodes), MOS capacitors and MOSFETs manufactured on high-resistivity p-type silicon before and after irradiation with gamma rays at doses between 10 Mrad(Si) and 500 Mrad(Si).

This work significantly extends the analyses and results presented in [7] in particular: i) by introducing a combined bulk and surface damage modeling scheme for real case LHC operation studies; ii) by extending the comparison between simulation findings and measurements with different and larger set of devices, irradiation doses and test structures; iii) by considering different operating conditions (e.g. operating temperatures and bias voltages) allowing the comparison with additional available literature data.

Manuscript received March 21, 2016; revised May 20, 2016.
This work has received funding from the European Union's Horizon 2020 research and innovation programme AIDA-2020 under grant agreement No 654168.

Francesco Moscatelli is with CNR - IMM Bologna, Via Gobetti 101, 40129 Bologna, Italy, e-mail: moscatelli@bo.imm.cnr.it (F. Moscatelli) and with INFN Perugia, via Pascoli, 06125 Perugia, Italy.
Daniele Passeri and Arianna Morozzi are with DI of University of Perugia and with INFN of Perugia, via Pascoli, 06125 Perugia, Italy.
Gian Mario Bilei is with INFN of Perugia, via Pascoli, 06125 Perugia, Italy.
Gian-Franco Dalla Betta and Roberto Mendicino are with DII of University of Trento and TIFPA-INFN, Via Sommarive 9,38123 Trento, Italy.

The overall aim of this study is to validate the suitability of the new TCAD modeling approach in view of its application for the design and optimization of the new generation of silicon detectors to be used in future HEP experiments.

## II. EXPERIMENTAL MEASUREMENTS

Test structures include MOS capacitors fabricated at the Institute for Microelectronics and Microsystems (IMM) of the Italian National Research Council of Bologna (Italy) on 4", *p*-type <100> silicon wafers with a nominal resistivity of 1Ω·cm. These MOS capacitors from IMM feature a 55 nm oxide thickness and they do not have a *p*-spray layer implanted. Additional test structures under study include gated-diodes, MOS capacitors and MOSFETs fabricated at FBK (Trento, Italy) on 6", *p*-type SiSi Direct Wafer Bonded Wafers from ICEMOS Technology Ltd [8]. The active layer is a Float Zone, <100> wafer, with a nominal resistivity higher than 3 kΩ·cm. The fabrication technology is "*n*-in-*p*", with a p-spray layer implanted at the surface to isolate adjacent n$^+$ collecting regions. The oxide thickness is 650 nm. For comparison purposes we eventually considered 15 test structures manufactured by FBK of Trento and 6 test structures with 4 MOS capacitors for each die manufactured by IMM of Bologna.

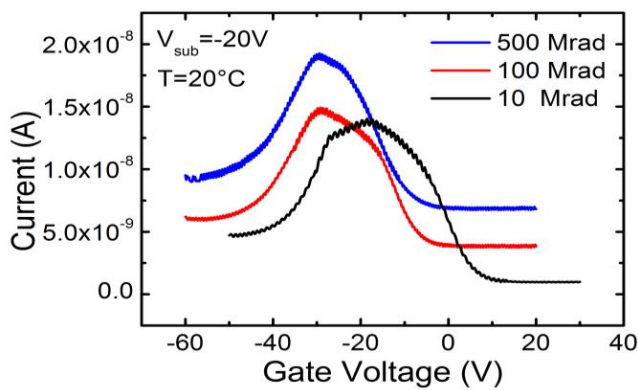

Fig. 1: Diode current as a function of gate voltage for interdigitated gated diodes [9] after gamma doses of 10 Mrad(Si), 100 Mrad(Si) and 500 Mrad(Si) and annealing at 80°C for 10 min.

The overall aim of this set of measurements was the extraction of the relevant parameters to be included within the TCAD simulation for the surface damage effect modeling. In particular, we looked at the flat-band voltage shift in order to evaluate the oxide charge density ($N_{OX}$). Moreover, the distribution of the interface traps ($D_{IT}$) as well as the density of the interface traps ($N_{IT}$) have been inferred from high-frequency and quasi-static capacitance measurements, as detailed below.

Current-Voltage (I-V) and Capacitance-Voltage (C-V) measurements have been carried out at 20°C using two computer-controlled parametric systems. The first one is based on a semiautomatic probe station (Micromanipulator P200A) with a HC-1000 thermal chuck system, and an Agilent B1500A semiconductor device analyzer. The second is based on a semi-automatic probe station (Micromanipulator MM6620) with a Temptronic TP315B thermo chuck, a Keithley K707 switching matrix and a KeithleyK90 I-V measuring system. Capacitance–Voltage (C–V) measurements were performed with a Keithley K82 system. Using High-Frequency (HF) and Quasi-Static (QS) C–V characteristics measured on MOS capacitors and I-V measurements on gated diodes, the oxide-charge density, the surface generation velocity ($s_0$) and the interface-trap density near the valence band have been determined before and after irradiation with gamma rays. From the gated diode current as a function of the gate voltage we calculated $s_0$ values of (0.54±0.07) cm/s before irradiation for all the measured test structures. For the C-V characterization of MOS capacitors the HF measurements were carried out at 100 kHz with a small signal amplitude of 15 mV. The QS characteristics were measured with delay times of 0.7 s using a voltage step of 100 mV. These measurements enabled us to obtain the total oxide charge ($Q_{OX}$), that includes the oxide fixed charge ($Q_{fix}$), interface charge ($Q_{it}$) and oxide trapped charge ($Q_{trap}$), considering for our processes the mobile ion charge ($Q_{mo}$) negligible. The interface state density was estimated by using the standard High-Low method described in [10].

Before irradiation the oxide charge density was of the order of $2 \times 10^{10}$ cm$^{-2}$, as extrapolated from flat-band voltages of about -1 V. The interface trap density is of the order of $10^9$-$10^{10}$ cm$^{-2}$ eV$^{-1}$ in the range 0.2-0.6 eV from the valence band edge. Gamma-ray irradiations have been performed without bias polarization on the same samples at the Gamma Irradiation Facility of Sandia National Laboratories (USA) with doses ranging from 10 to 500 Mrad(Si). Measurements have been carried out after an annealing at 80°C for 10 min to compare the results with [3, 4].

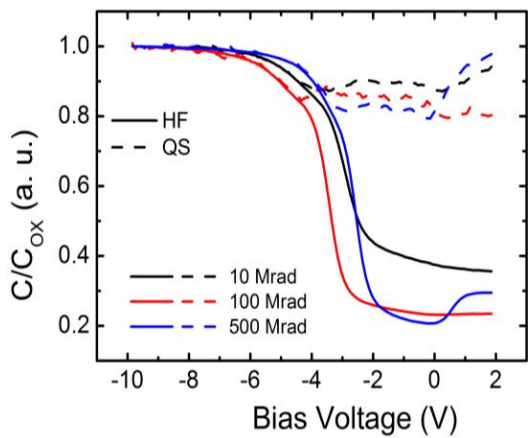

Fig. 2 High-frequency and quasi-static capacitance measured on MOS capacitors after gamma doses of 10 Mrad(Si), 100 Mrad(Si) and 500 Mrad(Si) and annealing at 80°C for 10 min.

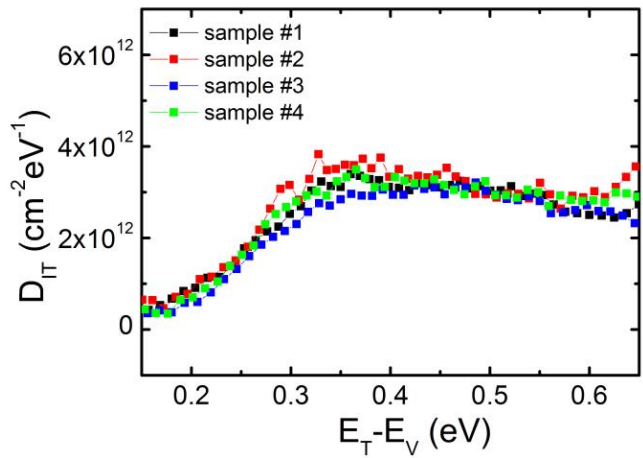

Fig. 3: Interface state density ($D_{it}$) near the valence band measured on four MOS capacitors after a gamma dose of 10 Mrad(Si) and an annealing at 80°C for 10 min.

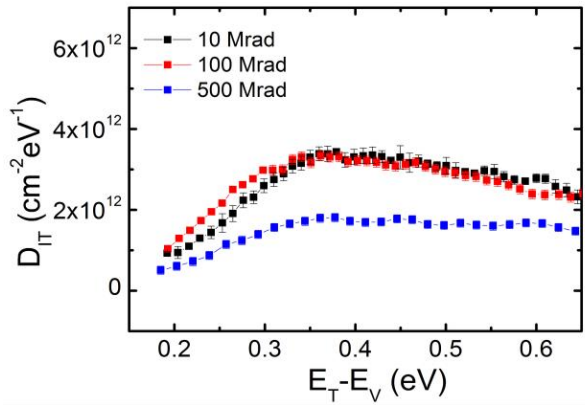

Fig. 4: Avergae interface state density ($D_{it}$) near the valence band measured on four MOS capacitors after a gamma dose of 10 Mrad(Si), 100 Mrad(Si) and 500 Mrad(Si) and annealing at 80°C for 10 min. The error bars represent standard deviation.

After irradiation, the surface generation velocity measured at 20 V, calculated from the diode current as a function of the gate voltage (Fig. 1), is of the order of (530±30) cm/s for all the measured test structures within the full range of irradiation doses (10-500 Mrad(Si)). The surface current density is of the order of 1.1-1.2 µA/cm² for all the gated diodes. The integrated density of interface traps related to this surface velocity has been estimated as $1\times10^{12}$ cm$^{-2}$ following the procedure reported in [11].

From the measurements on the MOS capacitors after irradiations we extrapolated flat-band voltages of about -3.5 V (Fig. 2) corresponding to an oxide charge density after 10 and 100 Mrad(Si) of the order of $1.2\times10^{12}$ cm$^{-2}$. After 500 Mrad(Si) the flat-band voltage is about -2.5V thus indicating an oxide charge density of about $0.8\times10^{12}$ cm$^{-2}$.

The high-low method allows us to extract the interface trap density as a function of the energy level near the valence band and the measured $D_{IT}$ after 10 Mrad(Si) on 4 MOS capacitors are shown in Fig. 3. The extracted interface trap density is of the order of $3\times10^{12}$ cm$^{-2}$ eV$^{-1}$ in the range 0.3-0.6 eV from the valence band after 10 and 100 Mrad(Si), while it is of the order of $1.5$-$2\times10^{12}$ cm$^{-2}$ eV$^{-1}$ in the range 0.3-0.6 eV from the valence band after 500 Mrad(Si) (Fig. 4). Integrating the interface trap density in the range 0.2-0.6 eV we obtain an integrated density of interface traps, which is of the order of $1\times10^{12}$ cm$^{-2}$ after 10 or 100 Mrad(Si) and $6\times10^{11}$ cm$^{-2}$ after 500 Mrad(Si) (Fig. 5).

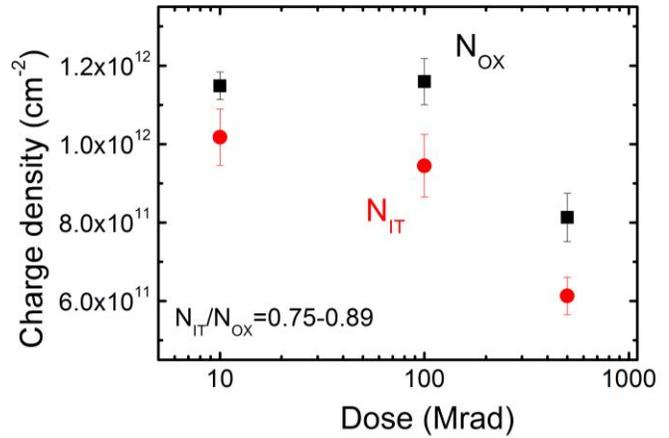

Fig. 5: The oxide charge density ($N_{OX}$) and the integrated density of interface traps ($N_{it}$) as a function of the gamma dose measured on MOS capacitors after an annealing at 80°C for 10 min.

The unexpected lower value of $N_{OX}$ and $N_{IT}$ at the highest dose (500 Mrad(Si)) has been confirmed by measurements carried out on different samples. This effect could be due to combined interface and bulk damage which can occur at very high doses with gamma irradiation [12] and it is still under investigation.

Integrated density of interface traps and oxide charge density extracted from C–V characteristics and current-voltage (I-V) measurements are summarized in Table I. Looking at a suitable modeling description of the interface trap state energies, an equivalent (de-convoluted) distribution of densities can be inferred. In particular, by analyzing the results of Fig. 3 it is possible to reproduce the measured spectral density distribution by a convolution of two Gaussian curves picked at $E_T=E_V+0.35$ eV and at $E_T=E_V+0.56$ eV, with a peak concentration of $2.5\times10^{12}$ cm$^{-2}$ eV$^{-1}$ and $2\times10^{12}$ cm$^{-2}$ eV$^{-1}$, respectively (Fig. 6).

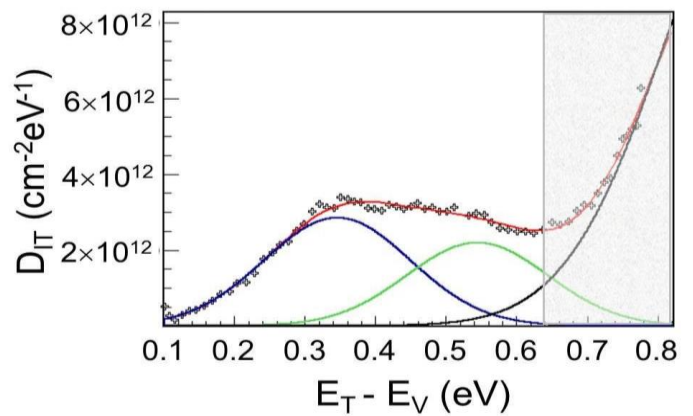

Fig. 6 Deconvolution of interface state density ($D_{it}$) near the valence band measured on MOS capacitors after a gamma dose of 10 Mrad(Si) and an annealing at 80°C for 10 min.

TABLE I
SURFACE CURRENT DENSITY (J) EXTRACTED FROM I-V CHARACTERISTICS OF GATED DIODES AND INTEGRATED DENSITY OF INTERFACE TRAPS ($N_{IT}$) AND OXIDE CHARGE DENSITY ($N_{OX}$) EXTRACTED FROM C–V CHARACTERISTICS MEASUREMENTS OF MOS CAPACITORS

|  | 10 Mrad(Si) | 100 Mrad(Si) | 500 Mrad(Si) |
|---|---|---|---|
| J (µA/cm$^2$) GCD | 1.33±0.08 | 1.11±0.07 | 1.17±0.08 |
| $N_{OX} \times 10^{12}$ (cm$^{-2}$) MOS | 1.15±0.04 | 1.16±0.06 | 0.81±0.06 |
| $N_{IT} \times 10^{12}$ (cm$^{-2}$) MOS | 1.02±0.07 | 0.95±0.08 | 0.61±0.05 |
| $D_{IT}$ (cm$^{-2}$ eV$^{-1}$) MOS | 2.5–3×10$^{12}$ (0.3–0.6 eV) | 2.5–3×10$^{12}$ (0.3–0.6 eV) | 1.4–1.8 ×10$^{12}$ (0.3–0.6 eV) |

## III. SIMULATIONS

### A. Surface damage model

To evaluate the surface effect in terms of strip isolation, a simple two strip structure featuring double p-stop structure (with low peak doping concentration), 4 micrometers wide and separated by 6 micrometers has been considered. The depth of the p-stop implant was 1.5 micrometers. The strip pitch was 80 micrometers. The n+ strip width was 18 micrometers. The substrate doping of the device was $3\times10^{12}$ cm$^{-3}$; the thickness of the device was set to 320 micrometers, accounting for a backside deep diffused p+ doped region of 30 micrometers.

For a comprehensive and physically sound interface traps description, we considered the combined effect of the donor interface traps described within Section II and two acceptor interface traps at $E_T=E_C-0.4$ eV and at $E_T=E_C-0.6$ eV following the findings in [3, 4]. The energy level of the donor trap was varied at different doses, and the effect of the energy level is shown in Fig. 7, eventually resulting in a model with the two acceptors and one donor level at $E_V+0.6$ eV. This donor level was extracted from measurements after gamma irradiation .

The interface trap density ($N_{IT}$) has been therefore set according to the measured oxide charge ($N_{OX}$) extracted from MOS capacitors. Moreover, for any given acceptor concentration $N_{IT}$, 60% of acceptors traps were allocated at $E_T=E_C-0.6$ eV, while the remaining 40% were allocated at $E_T=E_C-0.4$ eV according to [3, 4]. The capture cross sections of all defects were fixed at $1\times10^{-15}$ cm$^{-2}$. We conservatively used the same $N_{IT}$ for acceptor and donor traps while varying the $N_{IT}/N_{OX}$ ratio from 0.5 to 1. While the density of interface traps can not be extracted from measurement exactly, it has strong effect on the isolation properties of the device. We therefore looked closely at the interstrip resistance for different oxide charge values ($N_{OX}$) and trap state densities ($N_{IT}$).

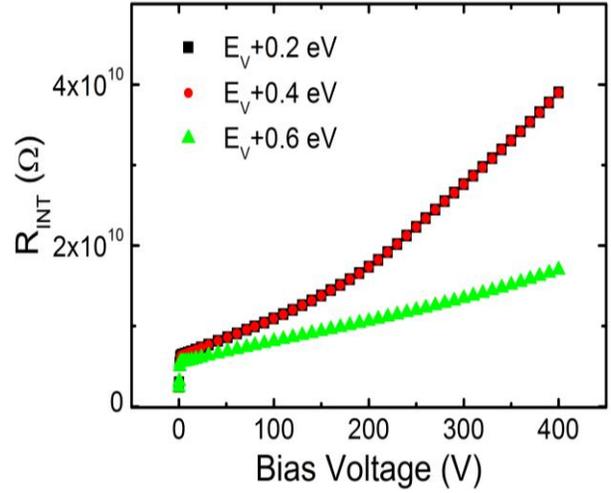

Fig.7 Simulated interstrip resistance as a function of the bias voltage at different energies of the donor interface traps . Simulations are obtained considering two acceptor interface traps at $E_T=E_C-0.4$ eV and at $E_T=E_C-0.6$ eV and one donor interface. Fluence $\Phi=1\times10^{15}$ cm$^{-2}$, $N_{IT}=1.6\times10^{12}$ cm$^{-2}$.

Experimental data from model validation purposes are taken from [13] for Hamamatsu Photonics (HPK) devices. In particular, Fig. 8-10 show the comparison between simulated and measured interstrip resistances as a function of bias voltage at different X-ray doses. The oxide charge densities are reported along the simulated curve for the different doses. It should be noticed that we have no exact information on the oxide charge build-up with fluence, which is strongly dependent on the technology options. However, a very realistic trend can be adopted for HPK technology, in particular being the highest $N_{OX}$ value close to the measured saturation value of $1.6$-$1.8\times10^{12}$ cm$^{-2}$ in the dose range 10-1000 Mrad(Si) [3].

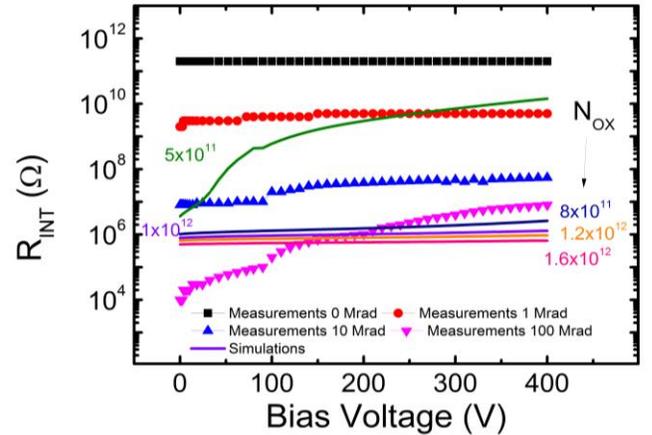

Fig.8 Simulated and measured [13] interstrip resistance as a function of $V_{BIAS}$ at different x-ray doses. Simulations are obtained considering two acceptor interface traps at $E_T=E_C-0.4$ eV and at $E_T=E_C-0.6$ eV and one donor interface trap at $E_V+0.6$ eV. The coloured numbers are related to the oxide charge density ($N_{OX}$). In this case $N_{IT}=0.5\times N_{OX}$.

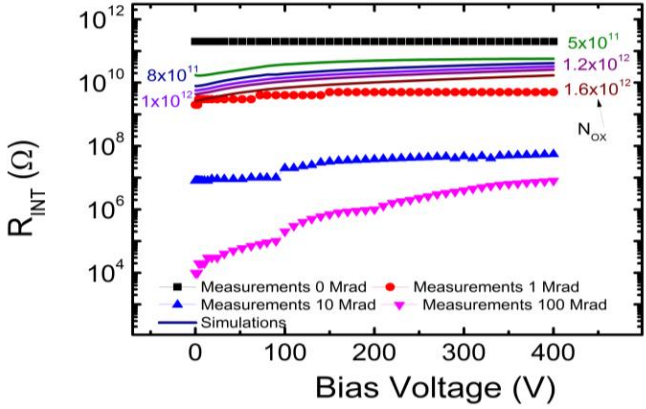

Fig.9 Measured [13] and simulated interstrip resistance as a function of $V_{BIAS}$ at different x-ray doses. Simulations are obtained considering two acceptor interface traps at $E_T=E_C-0.4$ eV and at $E_T=E_C-0.6$ eV and one donor interface trap at $Ev+0.6$ eV. The coloured numbers are related to the oxide charge density ($N_{OX}$). In this case $N_{IT}= N_{OX}$.

TABLE II
OXIDE CHARGE AND INTERFACE TRAP DENSITY INTRODUCED IN THE SURFACE DAMAGE MODEL CONSIDERING LITERATURE DATA PUBLISHED IN [3]

| Interface Defect | Level | Concentration |
|---|---|---|
| Acceptor | $E_C-0.4$ eV | 40% of acceptor $N_{IT}$ ($N_{IT}=0.8 \cdot N_{OX}$) |
| Acceptor | $E_C-0.6$ eV | 60% of acceptor $N_{IT}$ ($N_{IT}=0.8 \cdot N_{OX}$) |
| Donor | $Ev+0.6$ eV | 100% of donor $N_{IT}$ ($N_{IT}=0.8 \cdot N_{OX}$) |

Within this framework, a key role is played by the interface trap state density as well; we therefore considered an extensive analysis using different value of $N_{IT}$. To begin with, a relatively low trap density has been considered, in particular for the case $N_{IT}/N_{OX}=0.5$ (Fig. 8). As can be seen from the comparison, with this $N_{IT}/N_{OX}$ ratio it is not possible to reproduce correctly the interstrip resistance as a function of the dose. The predicted isolation behavior is actually too low for values of $N_{OX}$ of the order of $1\times10^{12}$ cm$^{-2}$.

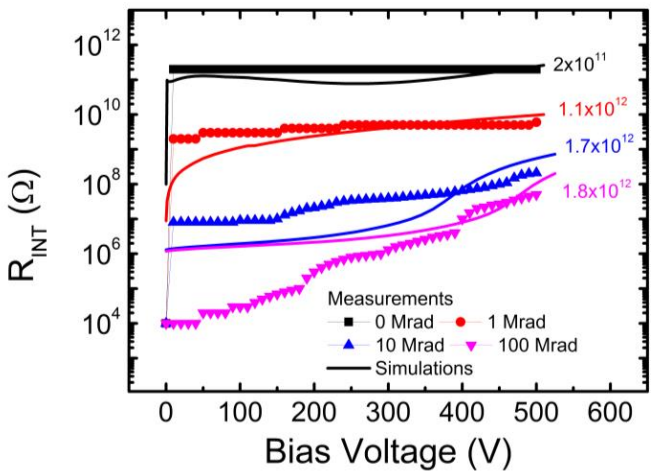

Fig.10 Measured [13] and simulated interstrip resistance as a function of $V_{BIAS}$ at different x-ray doses. Simulations are obtained considering two acceptor interface traps at $E_T=E_C-0.4$ eV and at $E_T=E_C-0.6$ eV and one donor interface trap at $Ev+0.6$ eV. In this case $N_{IT}=0.8\times N_{OX}$.

Fig. 9 shows the same comparison for $N_{IT}/N_{OX}=1$. In this case, a dual behavior has been found: the interstrip resistance is always very high even at very high doses (namely, corresponding to very high values of $N_{OX}$), being the strips always isolated.

Finally, Fig. 10 shows the comparison of measured and simulated interstrip resistance as a function of bias voltage at different X-ray doses for the case $N_{IT}/N_{OX}=0.8$. By consistently increasing the oxide charge and consequently the interface trap densities with the X-ray dose, it is possible to match the experimental data using the parametrization described previously and summarized in Table II. In particular within the high voltage range (i.e. the operating region of interest) the isolated (unirradiated) condition is well reproduced ($Q_{OX}=2.0\times10^{11}$), as well as the worst condition of isolation loss due to the irradiation ($Q_{OX}=1.8\times10^{12}$).

### B. Comprehensive Surface and Bulk damage model

At the very high fluences expected at HL-LHC (e.g. greater than $1\times10^{16}$ n/cm$^2$) and at the expected operating voltages, bulk damage effects have still a dominant role in setting the charge collection efficiency, depletion voltage and leakage current while oxide charge and interface traps build-up have only a minor effect on such parameters. On the other hand, oxide charge and interface traps play a dominant role in setting the isolation properties of adjacent strips. This is of utmost importance for segmented detectors (microstrip or pixel detectors).

Once assessed the main parameters of the surface damage effects to be included within the TCAD model, a comprehensive TCAD model has been devised by combining the bulk damage model [14] and the surface model previously described. A comprehensive analysis of the variation of the effective doping concentration (Neff), the leakage current density and the charge collection efficiency (CCE) as a function of the fluence has been performed using the Synopsys Sentaurus TCAD device simulator. As a reminder, the bulk trap level parameters used here are designed to model p-type Float Zone substrates after irradiation, and are based on work done by our group in the past [1,14]. The full details of the traps are given in Table III, IV, V. Each defect level is characterized by different parameters: the energy level (eV), the associated defect type, the cross sections for electrons $\sigma_e$ and holes $\sigma_h$ (cm$^{-2}$) and the introduction rate $\eta$ (cm$^{-1}$), respectively.

For the purpose of evaluating the variation in depletion voltage ($V_{FD}$) and leakage current ($I$), a 280 μm-thick n-in-p pad detector was simulated at different fluences. Capacitance-Voltage characteristics were used to calculate the full depletion voltage of the detectors under study. The standard procedure used for the extraction of $V_{FD}$ was the fit with two straight lines in the logC-logV plot near the kink [15].

The choice of the kink point in the case of the irradiated detectors has some degree of uncertainty due to the non-constant value of the capacitance at full depletion. The reference measurements taken for the comparison with simulation findings are extracted from [16]. The resulting depletion voltages in Fig.11 show indeed a good agreement, within the experimental errors, between simulations and measurements at room temperature. The OldSlotboom bandgap narrowing model has been used for the computation of the band edge references [17].

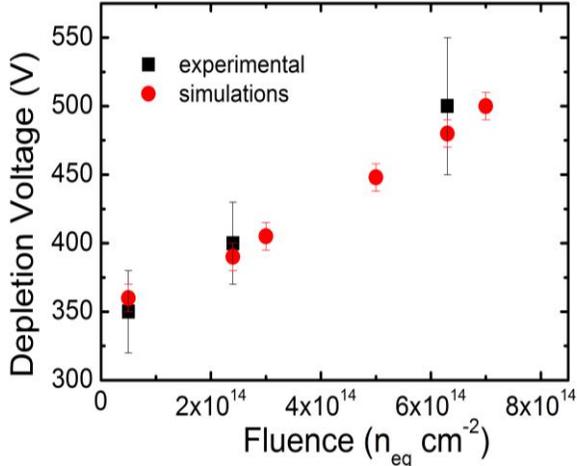

Fig. 11 Comparison between simulated and experimental [16] depletion voltages for n-in-p pad detectors at room temperature.

TABLE III
THE RADIATION DAMAGE MODEL FOR P-TYPE
(UP TO $7\times10^{15}$ N/CM$^2$)

| Type | Energy (eV) | $\sigma_e$ (cm$^{-2}$) | $\sigma_h$ (cm$^{-2}$) | $\eta$ (cm$^{-1}$) |
|---|---|---|---|---|
| Acceptor | Ec-0.42 | $1\times10^{-15}$ | $1\times10^{-14}$ | 1.613 |
| Acceptor | Ec-0.46 | $7\times10^{-15}$ | $7\times10^{-14}$ | 0.9 |
| Donor | Ev+0.36 | $3.23\times10^{-13}$ | $3.23\times10^{-14}$ | 0.9 |

TABLE IV
THE RADIATION DAMAGE MODEL FOR P-TYPE
(IN THE RANGE $7\times10^{15}$-$1.5\times10^{16}$ N/CM$^2$)

| Type | Energy (eV) | $\sigma_e$ (cm$^{-2}$) | $\sigma_h$ (cm$^{-2}$) | $\eta$ (cm$^{-1}$) |
|---|---|---|---|---|
| Acceptor | Ec-0.42 | $1\times10^{-15}$ | $1\times10^{-14}$ | 1.613 |
| Acceptor | Ec-0.46 | $3\times10^{-15}$ | $3\times10^{-14}$ | 0.9 |
| Donor | Ev+0.36 | $3.23\times10^{-13}$ | $3.23\times10^{-14}$ | 0.9 |

TABLE V
THE RADIATION DAMAGE MODEL FOR P-TYPE
(IN THE RANGE $1.6\times10^{16}$-$2.2\times10^{16}$ N/CM$^2$)

| Type | Energy (eV) | $\sigma_e$ (cm$^{-2}$) | $\sigma_h$ (cm$^{-2}$) | $\eta$ (cm$^{-1}$) |
|---|---|---|---|---|
| Acceptor | Ec-0.42 | $1\times10^{-15}$ | $1\times10^{-14}$ | 1.613 |
| Acceptor | Ec-0.46 | $1.5\times10^{-15}$ | $1.5\times10^{-14}$ | 0.9 |
| Donor | Ev+0.36 | $3.23\times10^{-13}$ | $3.23\times10^{-14}$ | 0.9 |

As a further validation of the simulation scheme, the increase of the leakage current with fluence has been evaluated as well. Experimentally, the leakage current of damaged silicon device increases with temperature. Therefore, for sake of comparison, leakage current values are normalized to 293K. We considered the Hurkx Trap Assisted Tunneling model [18] to consistently take into account the effect of trap states on the leakage current variation after depletion. Fig. 12 shows how the simulation predicts the increase of the leakage current at room temperature with the fluence and the saturation of the current at full depletion voltage.

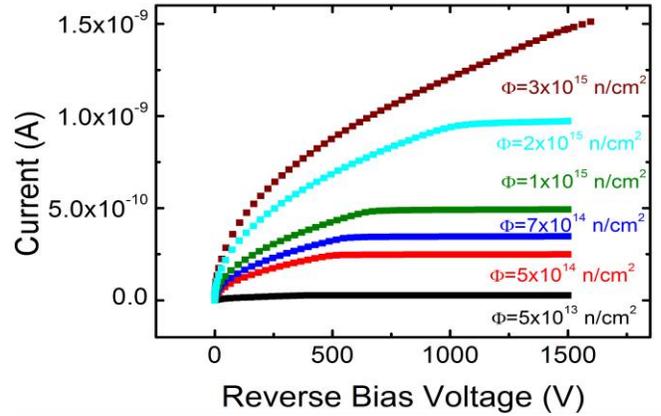

Fig. 12 Simulated leakage current as a function of the applied reverse voltage at different fluences for p-type silicon substrates.

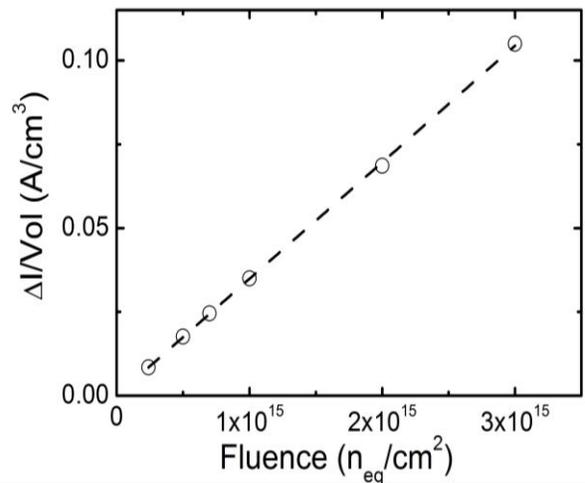

Fig. 13 Simulation of the leakage current density increase at 293 K as a function of the fluence for p-type detectors. The slope of the linear fitting is the radiation damage constant $\alpha$, which is $(3.48\pm0.03)\times10^{-17}$ A/cm. The R-squared of the fitting is 0.999.

Since the leakage current measurement uncertainty can limit the generality of a single set and/or specific sample data vs. simulation comparison, we decide to use, as a figure of merit of the simulation modelling, the radiation damage constant defined by:

$$\alpha = \frac{I}{\Phi_{eq} \cdot Vol}.$$

where *Vol* is the volume of the device. In particular, we considered the increase of the leakage current (*I*) at increasing fluences ($\Phi_{eq}$) at a given $V_{BIAS}$ after the full depletion of the device. Thanks to the virtual framework, we can extend the calculation of $\alpha$ up to very high fluences. However, since the required bias voltage to fully deplete the device would become unrealistically high, we limited the reported $\alpha$ values to a fluence of $3\times10^{15}$ n/cm$^2$.

The simulation findings give a value α = (3.48±0.03)×10⁻¹⁷ A/cm (Fig. 13), which is close (around 12%) to the experimental value of α=(3.99±0.03)×10⁻¹⁷ A/cm measured at 293K after an 80 min anneal at 60°C [19].

As a comprehensive validation of the modeling, the "active" behavior of a particle detector, i.e. the charge collection efficiency as a function of the fluence, has been investigated in detail.

The very high fluences and the operating voltages expected at the HL-LHC, make it necessary to consider the so called "multiplication" effect due to impact ionization. The avalanche generation which can be triggered at such high fluences has been considered by comparing different models available within Synopsys Sentaurus TCAD. Besides the default Van Overstraeten-De avalanche model [20], we considered Lackner [21], Okuto-Crowell [22], and University of Bologna [23] models. Different parameterization was used in order to evaluate their respective sensitivities. However, as the variation in charge collection efficiency between different models was within 3-4%, it was safe to use the default Van Overstraeten-De avalanche model

The charge collection behaviour of a single n+/p junction on 300 µm thick p-type substrate was simulated, at increasing fluences, spanning the whole range of values expected at HL-LHC. This is a simplified approach which can be used to set-up an equivalent, segmented detector. In order to be as much consistent as possible with measured structures, boundary effects at the lateral strip junctions have been considered. To this purpose, lateral diffusion of dopants as well as side oxide has been included within the simulated domain.

The charge collection efficiency (CCE) has been evaluated by considering a Minimum Ionizing Particle (MIP) hitting the sample diode perpendicularly to the front side contact. The transient behaviour of the device was simulated over time, and the current at the readout electrode was integrated over 20 ns, after subtracting the leakage current pedestal in order to find the total charge collected.

In Fig. 14 the comparison between simulated and experimental [24] charge collection in n-in-p strip detectors at 248 K and 900 V bias is reported for the full range of expected doses at HL-LHC. In this range of temperatures the variation of capture cross section is of the order of 5% [25], so we have considered constant capture cross section values. The oxide charge density has been considered saturated at $N_{OX}$=1.8×10¹² cm⁻² . In case of proton irradiation, this saturation value is reached at relatively low fluences of the order of 1×10¹⁴ 1-MeV n/cm², the exact value depending on the proton energy [26]. In case of neutrons, the total ionizing dose is due to the gamma-ray background, and oxide charge saturation is reached at fluences of the order of 1x10¹⁶ 1-MeV n/cm² [27]. Since surface damage has an effect on the charge collection efficiency only for very large fluences higher than 1x10¹⁶ 1-MeV n/cm², the oxide charge density can be assumed to be saturated for both protons and neutrons.

Surface damage has influence on the electric field, and this in turns significantly affect the charge collection properties, in particular at very high fluences, e.g. greater than 1×10¹⁶ n/cm².

In this case, an overestimation of the collected charge is inferred from the simulations (see inset in Fig. 14). This is due to the high field region at the strip corners which may act as avalanche generation preferred regions. The combined effects of (positive) oxide charge and interface traps tend to mitigate this effect, and cannot be neglected.

A very good agreement along the whole fluence range has been obtained, thus fostering the application of this model for the design and optimization of the new generation of silicon detectors to be used in the next generation of HEP experiments.

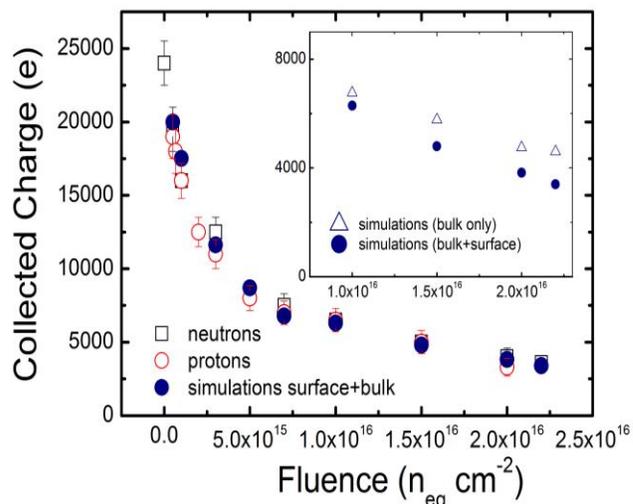

Fig.14 Comparison between simulated and experimental charge collection [24] in n-in-p strip detectors at 248 K and 900 V bias. In the inset the effect of the surface damage on the charge collection is shown for high fluences (for particle hit interesting the high-field regions).

IV. CONCLUSIONS

A new combined TCAD radiation damage modelling scheme, featuring both bulk and surface radiation damage effects, has been proposed and validated through the comparison of simulations and experimental measurements. In particular, a surface damage model has been developed by introducing the relevant parameters ($N_{OX}$, $N_{IT}$) extracted from experimental measurements carried out on p-type substrate test structures after gamma irradiations in the range 10-500 Mrad(Si). The combined surface and bulk model was able to reproduce the macroscopic effects of radiation damage in terms of depletion voltage and leakage current by considering impact ionization and deep-level cross-sections variation. Finally, the modeling scheme was able to reproduce the charge collection properties of devices irradiated up to fluences of the order of 2.2×10¹⁶ n/cm² This fosters the suitability of the model as a predictive tool for investigating sensor behavior at different fluences, temperatures, and bias voltages for the optimization of both 3D and planar silicon detectors for future HL-LHC High Energy Physics experiments.

V. ACKNOWLEDGMENTS

The authors gratefully acknowledge Sally Seidel and Martin Hoeferkamp of University of New Mexico (USA) for the support during the irradiation phases.